IOP Publishing          J. Phys. Commun. 3 (2019) 035003          https://doi.org/10.1088/2399-6528/ab07d1# Journal of Physics Communications

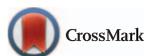

PAPER

OPEN ACCESS

RECEIVED
24 January 2019

ACCEPTED FOR PUBLICATION
18 February 2019

PUBLISHED
6 March 2019# Asymmetric localized states in periodic potentials with a domain-wall-like Kerr nonlinearity

Jincheng Shi[1,2,3] and Jianhua Zeng[1,2]

[1] State Key Laboratory of Transient Optics and Photonics, Xi'an Institute of Optics and Precision Mechanics of CAS, Xi'an 710119, People's Republic of China
[2] University of Chinese Academy of Sciences, Beijing 100084, People's Republic of China
[3] Key Laboratory for Physical Electronics and Devices of the Ministry of Education & Shaanxi Key Lab of Information Photonic Technique, Xi'an Jiaotong University, Xi'an 710049, People's Republic of China

E-mail: zengjh@opt.ac.cnOriginal content from this work may be used under the terms of the Creative Commons Attribution 3.0 licence.

Any further distribution of this work must maintain attribution to the author(s) and the title of the work, journal citation and DOI.

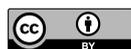Keywords: Bose–Einstein condensates in periodic potentials, solitons, photonic crystals, nonlinear optics at surfaces, fiber Bragg gratings, Kerr effect## Abstract

We study the existence of one-dimensional localized states supported by linear periodic potentials and a domain-wall-like Kerr nonlinearity. The model gives rise to several new types of asymmetric localized states, including single- and double-hump soliton profiles, and multihump structures. Exploiting the linear stability analysis and direct simulations, we prove that these localized states are exceptional stable in the respective finite band gaps. The model applies to Bose–Einstein condensates loaded onto optical lattices, and in optics with period potentials, e.g., the photonic crystals and optical waveguide arrays, thereby the predicted solutions can be implemented in the state-of-the-art experiments.## 1. Introduction

Periodic potentials (structures) are well-known and commonly used in physical science, among which undoubtedly the most familiar case is the crystal lattice in solids [1, 2]. In fact, we can say, without exaggeration, that the most intriguing feature of such periodic system is the emergence of forbidden band gaps where the waves pertaining to the corresponding wave spectra are not permitted to transmit. Because of the existence of resonant Bragg scattering from the periodic structures, such spectral gaps, in particular, give rise to a strong destructive interference of multiple reflections of waves, and accordingly, can greatly constrain the wave dispersion and diffraction in nature [3]. The control of wave dynamics in periodic potentials, during the last decades, has aroused growing interest of numerous researchers from different disciplines and fields, and significant progress has been made in exploring new and more complex periodic structures, such as the photonic crystal fibers/waveguides [1–3], optically induced photonic lattices [4–6], optical lattices (optical periodic potentials formed by counter-propagating laser beams) [7, 8], etc.

When the self-focusing or defocusing nonlinearity comes into play, the physical systems with periodic potentials show a lot of new phenomena [7–16]; most notably, besides the ordinary (fundamental) solitons supported by self-focusing effect, they can generate a novel class of solitons—the so-called bright gap solitons– the localized states residing inside the finite band gaps of the underlying linear spectrum under the action of defocusing nonlinearity and the strong Bragg scattering (which leads to anomalous dispersion because of the negative effective mass) described above, contrary to the well established concept that defocusing effect can only support dark solitons in uniform media [2]. It have demonstrated, over the past years, that the gap solitons can be created in diverse types of nonlinear periodic structures such as the optical fiber Bragg gratings [17], photonic crystals [2] and lattices [4–6], atomic Bose–Einstein condensates (BECs) trapped in optical lattices [14, 18, 19], and quite recently, the exciton-polariton BECs in semiconductor microcavities with reconfigurable lattice structures [20–22]. Further, other types of spatially localized states can be found in such nonlinear periodic settings as well, for example, vortex solitons carrying topological charge [23–25], multipole solitons [26] and truncated nonlinear Bloch waves (or gap waves) [27–31]. In particular, the latter case is self-trapped localized states with steep edges and arbitrary large atom numbers localized in a great number of deep optical lattice sites, which is distinct from the experimental realization of BEC gap solitons [19] under constraints of low atom numbers and densities.

© 2019 The Author(s). Published by IOP Publishing Ltd

In the meanwhile, periodic potentials have recently been extended to the nonlinearity coefficient, which are the 'nonlinear lattices' [32] what we now call. The inhomogeneous modulation of nonlinearity has widely used in nonlinear optics too, which is termed nonlinearity management [33]. Contemporary and cutting-edge fabrication technologies have made it possible to create structures with the nonlinearity landscape of unique properties featuring almost arbitrary transverse modulation [32]. As such, the study of soliton phenomena in truly periodic nonlinear lattice (the nonlinearity embedded into a linear uniform medium) is particularly intriguing, since it departs from the case in conventional linear lattices mentioned above with modulated refractive index. The stabilization of solitons and vortical ones in nonlinear lattices [34–43], combined linear and nonlinear lattices [44–47] is increasingly being studied in past years, and recently the interest is also on the scenarios with inhomogeneous modulations of nonlinearity [48–56]. In particular, spatially inhomogeneous nonlinear media with a defocusing nonlinearity, whose local strength grows fast enough from the pivot to the periphery, can uphold a vast variety of localized states, both the fundamental and higher-order solitons, which are in the forms of solitary vortices (with arbitrarily vortex charges) [48], vortex rings [48], soliton gyroscopes [49] and skyrmions [50], hopfions, complex hybrid modes, localized dark solitons and vortices [56].

In this work, we study in detail, theoretically and numerically, the existence and dynamics of one-dimensional matter wave localized states in an optical lattice with a domain-wall-like Kerr nonlinearity, characterized by different local (interatomic interaction) constant strengths at the two semi-infinite regions (and thus can be expressed by a step function). The model is able to support several new types of asymmetric localized states, including single- and double-hump soliton profiles, and multihump structures, which are stable in wide linear spectrum (finite band gaps) regions, verified by linear stability analysis and systematic simulations. We discuss the physical mechanism of the stabilization of these localized structures by such a combined linear lattice periodic potentials and a simple step-function nonlinearity model.

We stress that there is a very clear contrast between our model and the scenarios of semi-infinite periodic lattices (the cases with a interface between an optical lattice and uniform media, while keeping the nonlinearity constant), which were widely investigated for the surface gap solitons possessing a combination of the unique properties exhibited by gap solitons and common features typical for nonlinear surface waves [57–60]. Physically, such type of nonlinear surface solitons are spatially localized at the interface, since the chemical potential (or propagation constant, in optics), whose value is imaginary for a gap soliton residing at the optical lattice half interface, decides the penetration depth into uniform media, and therefore forming evanescent wave in this half [57]. Since the presence of exponentially decaying tail (evanescent wave) in the uniform half, the surface gap solitons existed merely at the linear interface [57], while in our model the nonlinear interface just tunes the amplitude and shape of solitons, we thus do not term them surface solitons. Our model is different from the purely nonlinear interfaces supported by two different nonlinear coefficients (while the linear potential is constant) [61–64], either [in particular, the theory of light-beam propagation at nonlinear interfaces had been well developed in the literatures [63, 64] three decades ago]. Because of lacking of linear lattices with spatially modulated refractive index and thus the fascinating tunable band gap in the system's linear spectrum, such nonlinear interfaces model, however, cannot support the localized states (both solitons and vortical ones) of gap types. The model proposed here therefore introduces a full linear periodic potential and shares the properties of purely nonlinear interfaces.

The rest of this work is organized as follows. In section 2, we present our model and analyze the band-gap diagram of Bloch modes in the linear periodic systems (the non-interacting BECs in optical lattices), the linear stability analysis for the spatially localized nonlinear modes based on eigenvalue problem is also introduced in this section. Several kinds of asymmetric localized states, including single- and double-hump solitons, and multihump structures, are predicted and studied systematically by numerical computation in section 3. Section 4 discusses the experimental conditions for realizing the predicted solutions, concludes with a summary and further extension.

## 2. Model

### 2.1. The model and band-gap spectrum of matter waves

We consider dynamics of atomic BECs loaded into an optical lattice with a domain-wall-type Kerr cubic nonlinearity described by the mean-field Gross-Pitaevskii equation (or nonlinear Schrödinger equation) for dimensionless scale of the wave function $\psi(x, t)$:

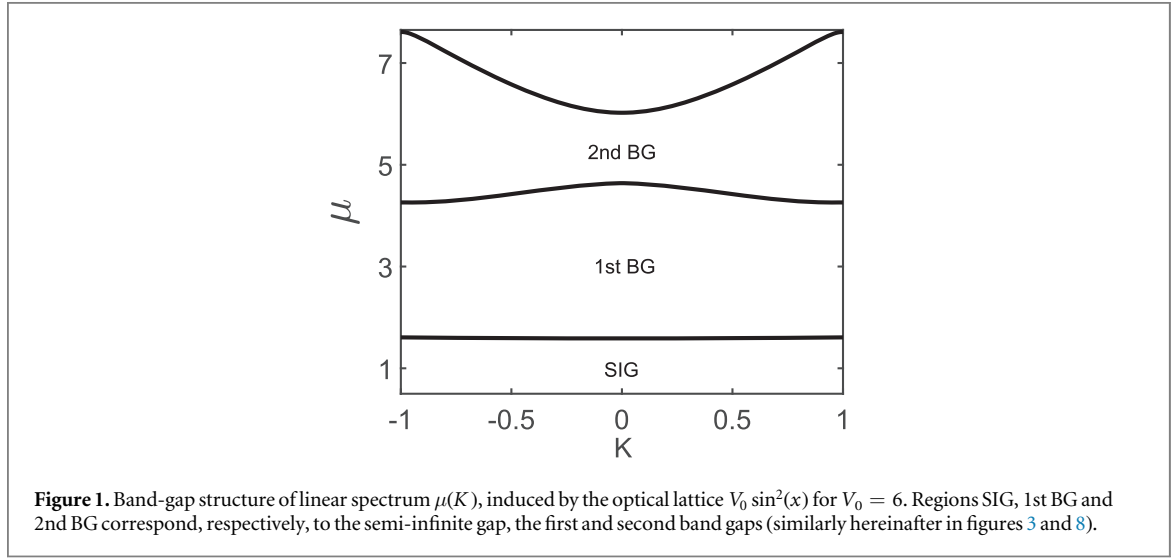

**Figure 1.** Band-gap structure of linear spectrum $\mu(K)$, induced by the optical lattice $V_0 \sin^2(x)$ for $V_0 = 6$. Regions SIG, 1st BG and 2nd BG correspond, respectively, to the semi-infinite gap, the first and second band gaps (similarly hereinafter in figures 3 and 8).

$$i\psi_t = -\frac{1}{2}\partial_x^2\psi + V_0 \sin^2(x)\psi + g(x)|\psi|^2\psi, \tag{1}$$

where $V_0$ is the strength of the optical lattice, and the nonlinear coefficient $g(x)$ yields:

$$g(x) = \begin{cases} g_l, & x \leqslant 0, \\ g_r, & x > 0. \end{cases} \tag{2}$$

Here the real constant coefficients $g_l$ and $g_r$ are in the same sign, and thus the nonlinearity suffers a sudden change only in magnitude. We define the parameter $\gamma = g_r/g_l$ for the sake of discussion, and set $|g_l| \equiv 1$. For comparison, we also discuss the uniform nonlinearity at two constant nonlinear coefficients $g_1$ and $g_2$ whose values will be specifically given below. The nonlinearity inherent to BECs because of the inevitable atom-atom collisions could be tuned with the popular technique called Feshbach resonance. We stress that the equation (1) also describes optical wave propagation in nonlinear optics with a replacement of time $t$ by propagation distance $z$.

The stationary solution of wave function $\psi(x, t)$ is usually sought as $\psi(x, t) = \phi(x)\exp(-i\mu t)$ with chemical potential $\mu$, in doing so the equation (1) satisfies

$$\mu\phi = -\frac{1}{2}\partial_x^2\phi + V_0 \sin^2(x)\phi + g(x)\phi^3. \tag{3}$$

To understand the matter-wave localized modes and their properties in the model with linear periodic lattice (optical lattice) it is necessary to give first the relevant band-gap structure. The linearization of equation (3) results into the following eigenvalue equation:

$$\mu\phi = -\frac{1}{2}\partial_x^2\phi + V_0 \sin^2(x)\phi. \tag{4}$$

The solution of eigenvalue problem (equation (4)) can produce the underlying band-gap spectrum $\mu(K)$, characterized by the momentum $K$ inside such periodic lattice. Specifically, according to the well-known Bloch's theorem borrowing from solid-state physics, eigenfunctions of equation (4) $\phi$ are periodic solutions known as Floquet-Bloch modes, and represented by their momentum $K$ inside the lattice, provided that eigenvalues $\mu$ inside the energy (Bloch) bands. Typical Floquet-Bloch mode at momentum $K$ is written as $\phi_K(x) = \Phi_K(x)\exp(iKx)$, here periodic function $\Phi_K(x)$ with a period equaling to that of the lattice. In doing so, we can get the band-gap spectrum for an optical lattice (we refer the readers to consult the very relevant papers in [65, 66] and books in [2, 3] for more details). As shown in figure 1 for the example at $V_0 = 6$, besides the usual semi-infinite gap, the first two finite band gaps are also existed. Inside these band gaps there are not any localized waves in linear case, as mentioned above, while such an acknowledgement would be overturned in the nonlinear scenario where a combination of periodicity and nonlinearity would lead to the appearance of families of different localized states in such linearly forbidden band gaps.

### 2.2. Stability analysis of localized modes

To settle the stability problem of localized modes, we take the perturbed solutions as

$$\psi = [\phi + \upsilon \exp(\lambda t) + i\omega \exp(\lambda t)]\exp(-i\mu t), \tag{5}$$

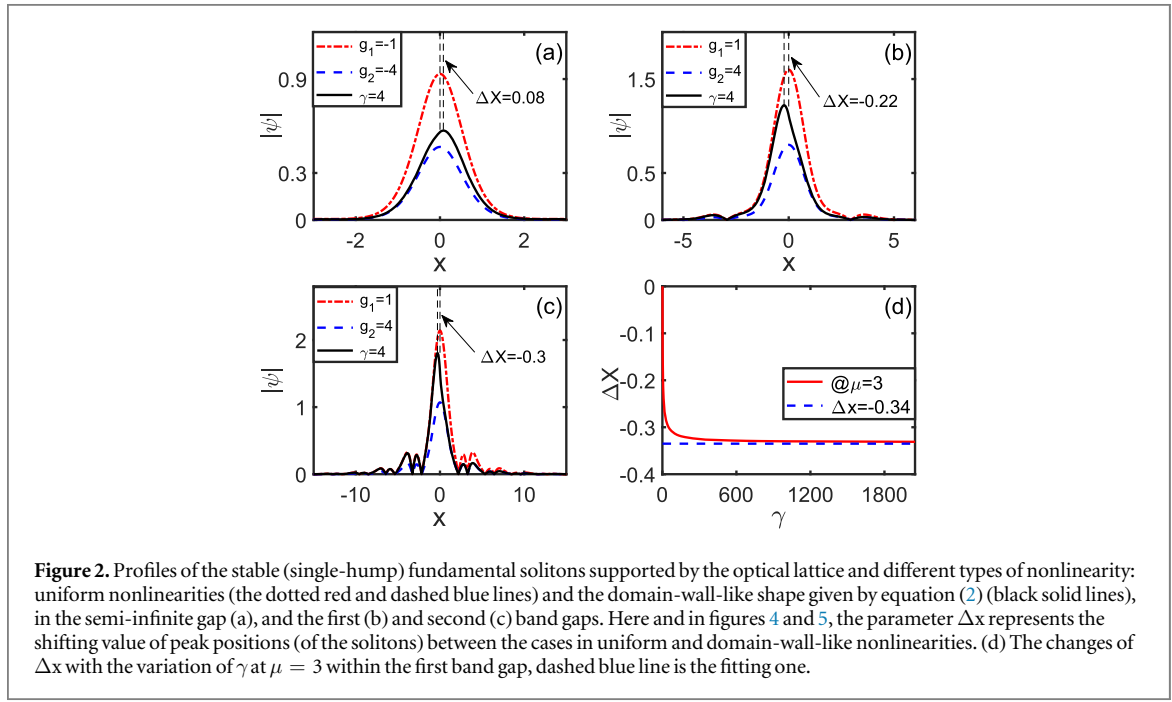

**Figure 2.** Profiles of the stable (single-hump) fundamental solitons supported by the optical lattice and different types of nonlinearity: uniform nonlinearities (the dotted red and dashed blue lines) and the domain-wall-like shape given by equation (2) (black solid lines), in the semi-infinite gap (a), and the first (b) and second (c) band gaps. Here and in figures 4 and 5, the parameter $\Delta x$ represents the shifting value of peak positions (of the solitons) between the cases in uniform and domain-wall-like nonlinearities. (d) The changes of $\Delta x$ with the variation of $\gamma$ at $\mu = 3$ within the first band gap, dashed blue line is the fitting one.

here $\upsilon$ and $\omega$ are the real and imaginary parts of infinitesimal perturbation eigenfunctions, $\lambda$ is the homologous perturbation eigenvalue (or growth rate). The linearization of equation (1) around stationary solution $\phi$ found from equation (3) results in the eigenvalue problem for $\lambda$

$$i\lambda \upsilon = -\frac{1}{2}\omega_{xx} + [\mu + V_0 \sin^2(x)]\omega + g(x)\phi^2\omega, \tag{6}$$

$$i\lambda \omega = -\frac{1}{2}\upsilon_{xx} + [\mu + V_0 \sin^2(x)]\upsilon + 3g(x)\phi^2\upsilon. \tag{7}$$

The linear stability analysis of perturbed localized solutions and associated growth rate $\lambda$ based on the perturbation equations (6) and (7) can be solved numerically, and evidently, the perturbed localized modes are stable as long as Re($\lambda$) = 0 for all the eigenvalues ($\lambda$).

## 3. Asymmetric spatially localized modes

### 3.1. Single- and double-hump matter-wave structures

We first investigate the formation of fundamental localized modes, e.g., the aforementioned single-hump wave structures in the forms of ordinary solitons and gap ones existing within semi-infinite gap and finite band gaps of the relevant linear spectrum, separatively, and under the self-attractive and self-repulsive nonlinearities. As a matter of fact, it should be noticed, in the condition of uniform nonlinearity, that such single-hump structures are known as symmetrical modes. For comparison, we have plotted these ordinary solitons in figure 2(a), and the gap solitons as populated in the first two band gaps in figures 2(b) and (c), for the two constant nonlinearities $g_1$ and $g_2$ (the former is in dotted red lines, and dashed blue lines for the latter), under the same chemical potential $\mu$. One can clearly observe that both kinds of solitons shrink at larger nonlinearity $g_2$, conforming to the nonlinear saturation mechanism of the physical system. And due to this fact, the corresponding single-hump localized wave structures, supported by the current domain-wall-like nonlinearity given in equation (2), stand in the middle of the cases with constant nonlinearities $g_1$ and $g_2$.

Interestingly, the unique property of our nonlinearity renders the localized waves stay away from the centre $x = 0$, resulting in asymmetrical shapes of the wave structures (besides the single-hump structures here in figures 2(a)–(c), asymmetric spatially localized modes are also for the double-hump structures and multihump ones below in figures 5 and 6). To elucidate the principle of this deviation, we define a parameter $\Delta x$ to measure the off-centre value between peak position of the solitons and geometric center $x = 0$. A scrutinized analysis found that the deviation $\Delta x > 0$ for ordinary solitons in figure 1(a), while $\Delta x < 0$ for localized modes inside the band gaps in figures 1(b), (c). The former case may be explained by the fact that, to maintain a balance in the current step nonlinearity, the asymmetric localized waves prefer to stay at the (right) side of larger self-focusing nonlinearity ($g_r = -4$) in which the threshold value of norm $N_0$ for generating a stable ordinary soliton is smaller compared to the other (left) side with lower nonlinearity ($g_l = -1$). For the latter in band gaps, by contrast, a bigger defocusing nonlinearity ($g_r = 4$), which repels the wave on the right half side, squeezes the wave to the left side ($g_l = 1$), leading to a negative $\Delta x$. Such deviation $\Delta x$ (absolute value) grows in much higher band gaps owning to their much stronger Bragg scattering, as seen from a comparison of figures 1(b) and (c). We further found, from numerous calculations, that the $\Delta x$ decreases quickly with an increase of $\gamma$, and then reaches to a certain value at large $\gamma$, as shown for the case at $\mu = -3$ (in first band gap) in figure 1(d).

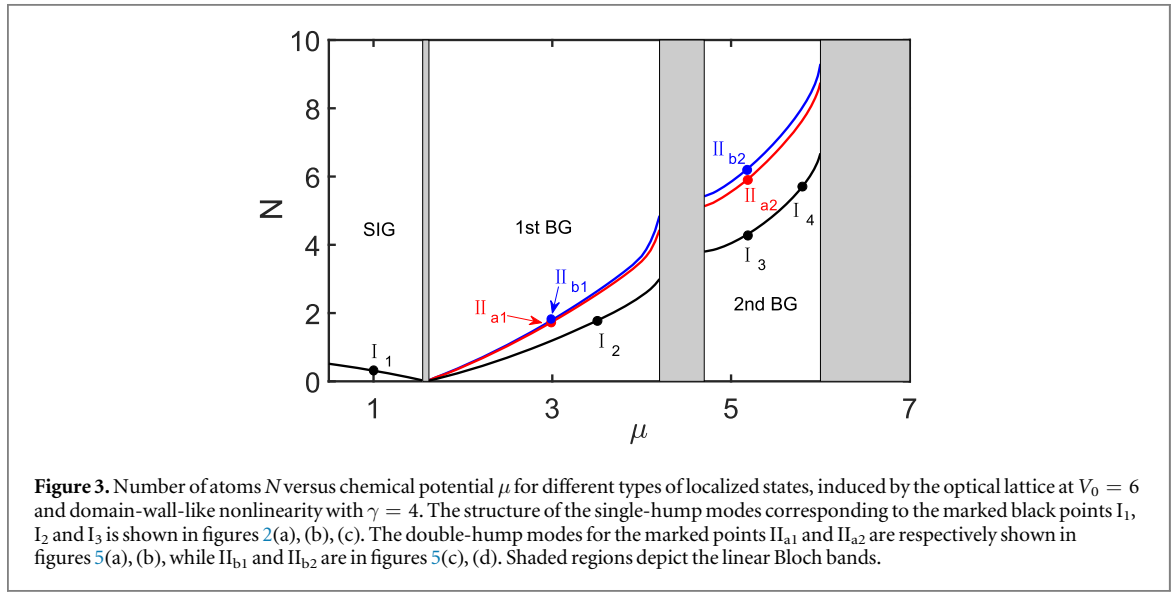

**Figure 3.** Number of atoms $N$ versus chemical potential $\mu$ for different types of localized states, induced by the optical lattice at $V_0 = 6$ and domain-wall-like nonlinearity with $\gamma = 4$. The structure of the single-hump modes corresponding to the marked black points $I_1$, $I_2$ and $I_3$ is shown in figures 2(a), (b), (c). The double-hump modes for the marked points $II_{a1}$ and $II_{a2}$ are respectively shown in figures 5(a), (b), while $II_{b1}$ and $II_{b2}$ are in figures 5(c), (d). Shaded regions depict the linear Bloch bands.

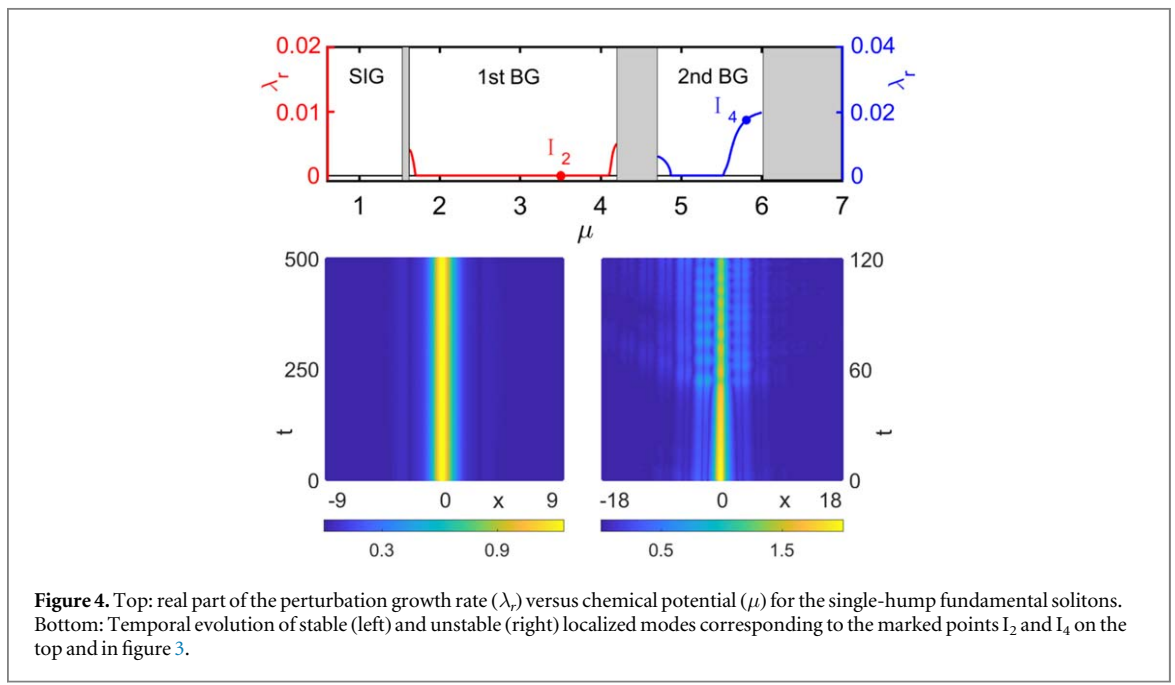

**Figure 4.** Top: real part of the perturbation growth rate ($\lambda_r$) versus chemical potential ($\mu$) for the single-hump fundamental solitons. Bottom: Temporal evolution of stable (left) and unstable (right) localized modes corresponding to the marked points $I_2$ and $I_4$ on the top and in figure 3.

A series of numerical computations, relied on the linear stability analysis (perturbation equations (6) and (7)) and dynamical equation (1), demonstrate that such single-hump matter-wave structures are very stable, exceptional cases are those near the band edges where exist weak oscillatory instability ($\lambda_r \sim 10^{-3} - 10^{-2}$) which, particularly, grows when going deeper inside band gaps, see the relevant linear stability results in the top panel of figure 4 and its representative dynamical evolution for both stable and unstable perturbation modes in the bottom left and right panels.

Besides the single-hump wave structures, the present model also supports different double-hump localized modes, which, depending on their shapes, can be categorized as two types—with one zero (node) and lack thereof, examples of them are plotted in figure 5. A common feature of both types localized modes is that they are composed by a single-hump localized state and a triple-hump one at constant nonlinearities $g_1$ and $g_2$, respectively. Their stability regions within the first two band gaps are collected in figure 3.

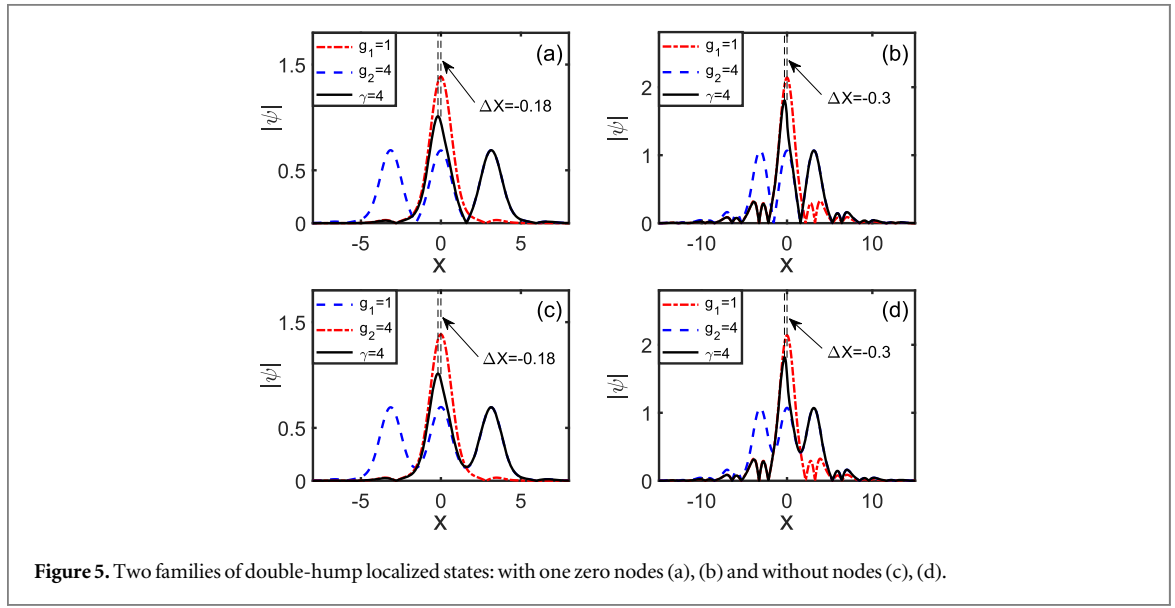

**Figure 5.** Two families of double-hump localized states: with one zero nodes (a), (b) and without nodes (c), (d).

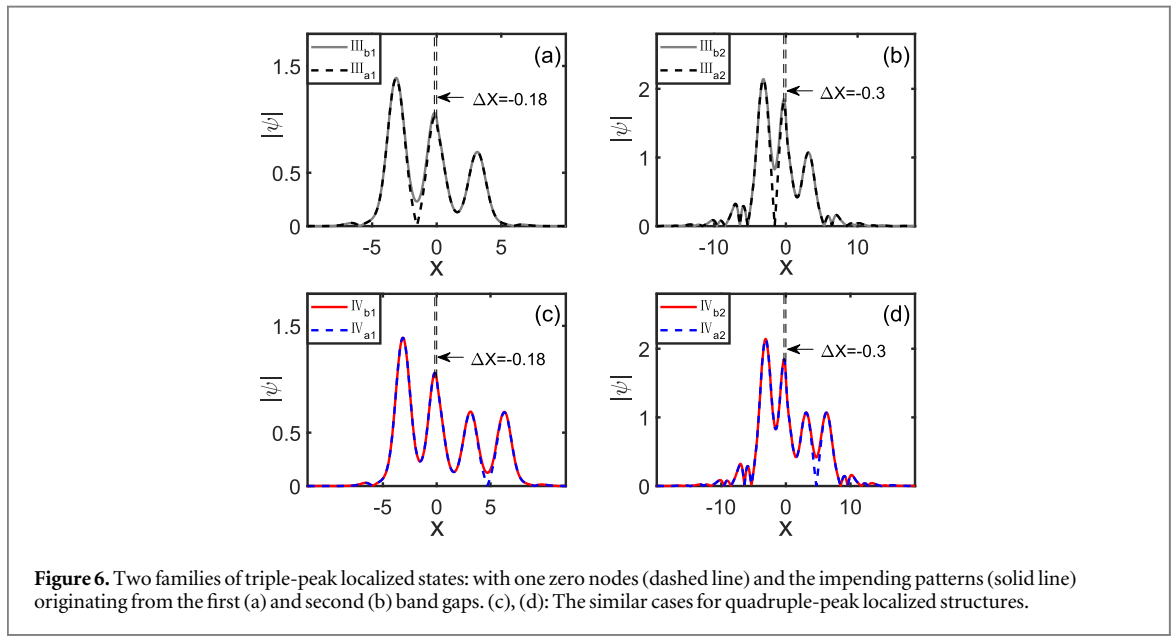

**Figure 6.** Two families of triple-peak localized states: with one zero nodes (dashed line) and the impending patterns (solid line) originating from the first (a) and second (b) band gaps. (c), (d): The similar cases for quadruple-peak localized structures.

Despite surface localized states, including surface solitons and gap ones, have been widely considered at two classes of interfaces: the purely nonlinear interface (or with an additional aperiodic potential) [61, 62], and at the interface between a semi-infinite lattice and uniform media (with a constant nonlinearity) [57–60], these models have restrictions: the former does not exhibit unique feature of linear periodic potential, the latter loses the nonlinearity-mediated (e.g., step nonlinearity) stabilization, and the lattice potential is only limited to half plate (not the whole space). Therefore, the investigation of localized states in the present model—full period potentials with a domain-wall-like nonlinearity—is relatively new. Our physical setting (and thereby the predicted asymmetric localized states) integrates specific properties typical for periodic potentials and nonlinear interfaces. It is also worthwhile mentioning here that the localized states, supported by the above two kinds of interfaces, are spatially localized at and near the interface, while a notable characteristic of the predicted asymmetric localized states (in our model) can be loosely localized and occupy many lattice sites (thus across the interface in both directions), see the following multiple-peak waves structures in figure 6.

### 3.2. Multiple-peak matter-waves structures

The above predicted localized states, including both single-hump fundamental solitons and double-hump ones, are all populated at and near the nonlinear interface, a natural question one may ask lies in whether localized states can be existed across such interface? If possible, subsequent questions arise: what are the formation conditions and how to generate them? Are they stable in finite band gaps? To this end, We now test the possibility for creating multiple-peak (or multihump) wave structures, which are of particular interest in experiments too, since the experimental observed BEC gap solitons were at very low atom numbers (it is thus a challenging task to realize them in conventional labs) [19].

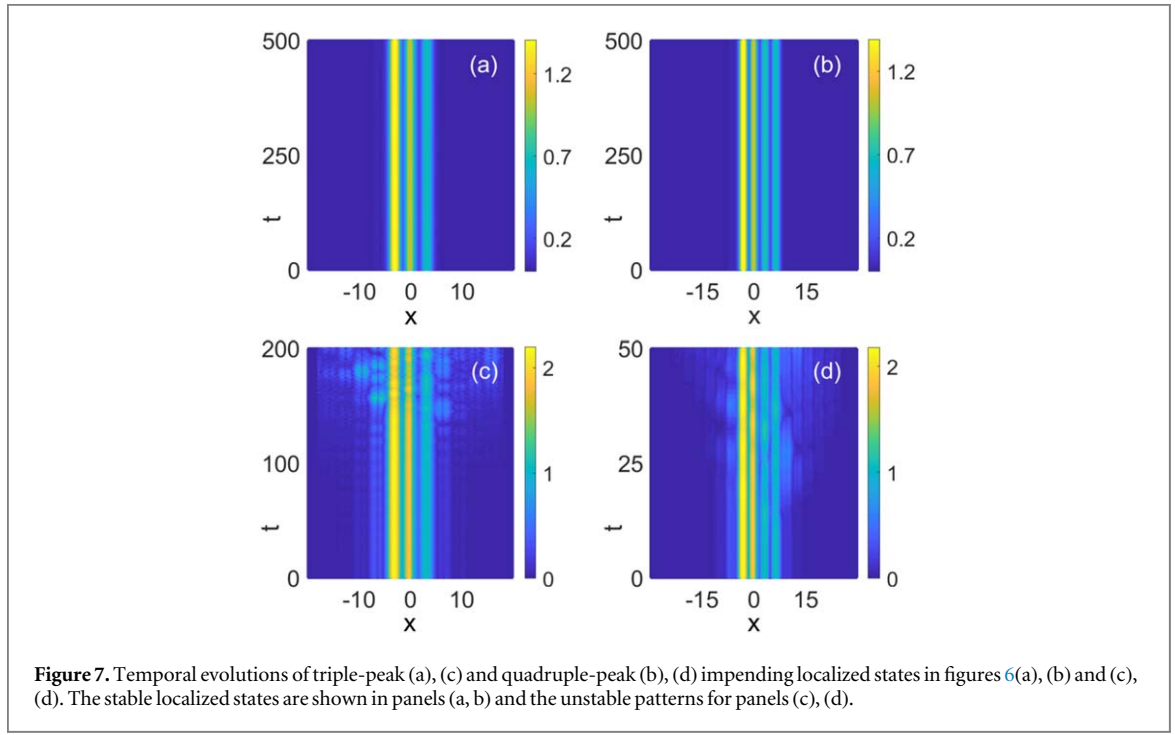

**Figure 7.** Temporal evolutions of triple-peak (a), (c) and quadruple-peak (b), (d) impending localized states in figures 6(a), (b) and (c), (d). The stable localized states are shown in panels (a, b) and the unstable patterns for panels (c), (d).

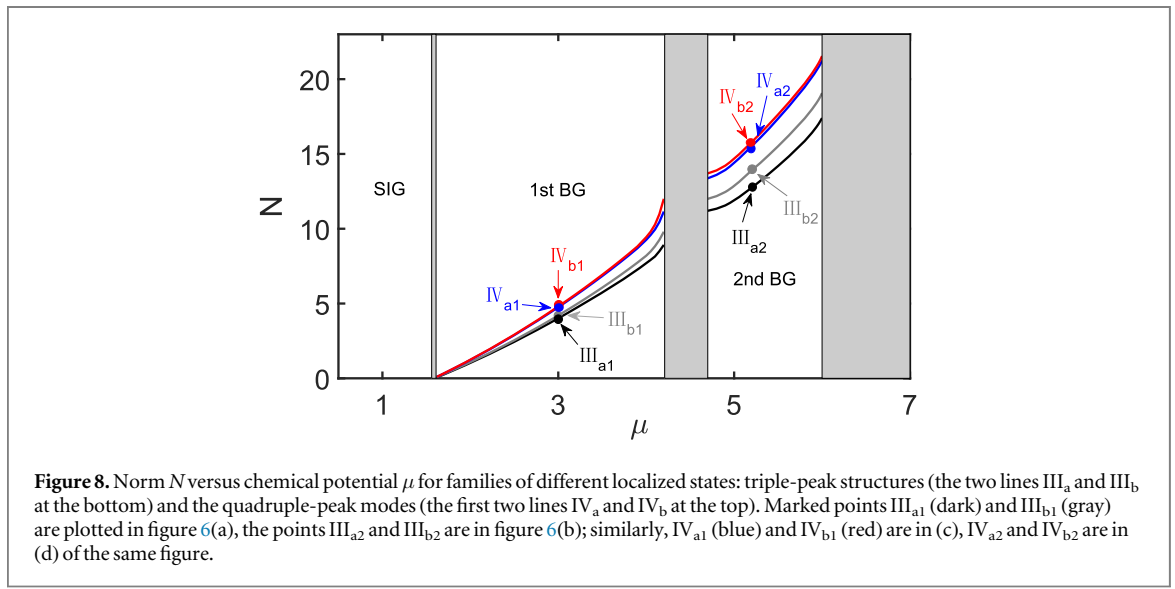

**Figure 8.** Norm $N$ versus chemical potential $\mu$ for families of different localized states: triple-peak structures (the two lines $III_a$ and $III_b$ at the bottom) and the quadruple-peak modes (the first two lines $IV_a$ and $IV_b$ at the top). Marked points $III_{a1}$ (dark) and $III_{b1}$ (gray) are plotted in figure 6(a), the points $III_{a2}$ and $III_{b2}$ are in figure 6(b); similarly, $IV_{a1}$ (blue) and $IV_{b1}$ (red) are in (c), $IV_{a2}$ and $IV_{b2}$ are in (d) of the same figure.

Numerical simulations suggest that the model also supports triple-peak and quadruple-peak localized states, as seen from the typical examples of the former lying in the first and second band gaps portrayed in figures 6(a) and (b). Representative modes for the latter are shown in the figures 6(c) and (d). We emphasize that both localized states occupy several lattice sites (hence they go beyond the nonlinear interface), and can as well be classified as two kinds depending on the number of zero (node)—with one and null, like their double-peak counterparts in figure 5. Stable evolutions of the triple-peak and quadruple-peak localized states are demonstrated in figure 7 through direct numerical simulations of them in real time. Their stability regions are collected in figure 8. Based on these results, we speculate and firmly believe that different matter-wave structures with more peaks—broad asymmetric localized states—can be constructed theoretically and realized in experiments.

## 4. Conclusion

We studied dynamics of one-dimensional atomic Bose–Einstein condensates (BECs) in an optical lattice with a domain-wall-like Kerr cubic nonlinearity, whose strength is made up of two different constants in both directions (the left and right half-planes), in the framework of the mean-field Gross-Pitaevskii equation. Several new types of localized states with asymmetric shape—asymmetric localized states, including single-, double- and multiple-hump modes, inside relevant gaps (both semi-infinite and the first two finite band gaps) of the underlying linear spectrum, are found. We have analyzed the physical mechanism of such localized wave structures, and addressed their stability properties by means of linear stability analysis and direct numerical simulations. A notable feature is the existence of stable multiple-peak wave structures in finite band gaps, making the observation of localized waves of gap type more accessible in conventional ultracold atoms laboratories, since a facing challenge to observe localized gap solitons is the low atom densities [the reported result is only about 250 atoms] [19].

Since Gross-Pitaevskii equation is the fundamental governing system equivalent to the nonlinear Schrödinger equation in nonlinear optics, thereby besides the BECs loaded into an optical lattice, the predicted solutions can also be realized in other periodic potentials, including fiber Bragg gratings, photonic crystals and lattice, as well as the ordinary waveguide arrays. As such, a natural extension is to study the asymmetric localized states and their propagation dynamics in such nonlinear optical systems, particularly in experiments. An issue of great interest is to consider more complicated situations with increasing dimensions and ingredients like the two-component system.

## Acknowledgments

This work was supported, in part, by the NSFC, China (project Nos. 61690224, 61690222), and by the Youth Innovation Promotion Association of the Chinese Academy of Sciences (project No. 2016357).

## ORCID iDs

Jincheng Shi  https://orcid.org/0000-0001-6928-1064
Jianhua Zeng  https://orcid.org/0000-0003-0189-1856

## References

[1] Pelinovsky D E 2011 *Localization in Periodic Potential: From Schrödinger Operators to the Gross-Pitaevskii equation* (Cambridge: Cambridge University Press)
[2] Kivshar Y S and Agrawal G P 2003 *Optical Solitons: From Fibers to Photonic Crystals* (San Diego, CA: Academic)
[3] Joannopoulos J D, Johnson S G, Winn J N and Meade R D 2008 *Photonic Crystals: Molding the Flow of Light* (Princeton: Princeton University Press)
[4] Christodoulides D N, Lederer F and Silberberg Y 2003 *Nature (London)* **422** 147
[5] Lederer F, Stegeman G I, Christodoulides D N, Assanto G, Segev M and Silberberg Y 2008 *Phys. Rep.* **463** 1
[6] Garanovich I L, Longhi S, Sukhorukov A A and Kivshar Y S 2012 *Phys. Rep.* **518** 1
[7] Brazhnyi V A and Konotop V V 2004 *Mod. Phys. Lett.* B **18** 627
[8] Morsch O and Oberthaler M 2006 *Rev. Mod. Phys.* **78** 179
[9] Malomed B A, Mihalache D, Wise F and Torner L 2005 *J. Optics B: Quant. Semicl. Opt.* **7** R53
[10] Kartashov Y V, Vysloukh V A and Torner L 2009 *Progress in Optics* ed E Wolf 52 (North Holland: Amsterdam) 63–148
[11] Malomed B A, Mihalache D, Wise F and Torner L 2005 *J. Optics B* **7** R53
[12] Baizakov B B, Malomed B A and Salerno M 2003 *Europhys. Lett.* **63** 642
Baizakov B B, Malomed B A and Salerno M 2004 *Phys. Rev.* A **70** 053613
[13] Yang J and Musslimani Z H 2003 *Opt. Lett.* **28** 2094
[14] Kevrekidis P G, Frantzeskakis D J and Carretero-González R (ed) 2008 *Emergent Nonlinear Phenomena in Bose–Einstein Condensates* (Berlin: Springer)
[15] Pethick C J and Smith H 2008 *Bose–Einstein Condensate in Dilute Gas* (Cambridge: Cambridge University Press)
[16] Carretero- González R et al (ed) 2013 *Localized Excitations in Nonlinear Complex Systems* (Heidelberg: Springer)
[17] Eggleton B J, Slusher R E, de Sterke C M, Krug P A and Sipe J E 1996 *Phys. Rev. Lett.* **76** 1627
[18] Ostrovskaya E A and Kivshar Y S 2004 *Phys. Rev. Lett.* **93** 160405
[19] Eiermann B, Anker T, Albiez M, Taglieber M, Treutlein P, Marzlin K-P and Oberthaler M K 2004 *Phys. Rev. Lett.* **92** 230401
[20] Ostrovskaya E A, Abdullaev J, Fraser M D, Desyatnikov A S and Kivshar Y S 2013 *Phys. Rev. Lett.* **110** 170407
[21] Cerda-Méndez E A, Sarkar D, Krizhanovskii D N, Gavrilov S S, Biermann K, Skolnick M S and Santos P V 2013 *Phys. Rev. Lett.* **111** 146401
[22] Tanese D et al 2013 *Nat. Commun.* **4** 1749
[23] Ostrovskaya E A and Kivshar Y S 2004 *Phys. Rev. Lett.* **93** 160405
[24] Sakaguchi H and Malomed B A 2009 *Phys. Rev.* A **79** 043606
[25] Zeng J and Malomed B A 2017 *Vortex Structures in Fluid Dynamic Problems* (Rijeka: InTech) ch 10 (https://doi.org/10.5772/66543)
[26] Lobanov V E, Kartashov Y V and Konotop V V 2014 *Phys. Rev. Lett.* **112** 180403

[27] Anker T, Albiez M, Gati R, Hunsmann S, Eiermann B, Trombettoni A and Oberthaler M K 2005 *Phys. Rev. Lett.* **94** 020403
[28] Alexander T J, Ostrovskaya E A and Kivshar Y S 2006 *Phys. Rev. Lett.* **96** 040401
[29] Zhang Y and Wu B 2009 *Phys. Rev. Lett.* **102** 093905
[30] Bennet F H, Alexander T J, Haslinger F, Mitchell A, Neshev D N and Kivshar Y S 2011 *Phys. Rev. Lett.* **106** 093901
[31] Bersch C, Onishchukov G and Peschel U 2012 *Phys. Rev. Lett.* **109** 093903
[32] Kartashov Y V, Malomed B A and Torner L 2011 *Rev. Mod. Phys.* **83** 247
[33] Malomed B A 2006 *Soliton Management in Periodic Systems* (Heidelberg: Springer)
[34] Sakaguchi H and Sakaguchi B A 2005 *Phys. Rev.* E **72** 046610
[35] Theocharis G, Schmelcher P, Kevrekidis P G and Frantzeskakis D J 2005 *Phys. Rev. A* **72** 033614
[36] Abdullaev F K and Garnier J 2005 *Phys. Rev. A* **72** 061605(R)
[37] Sivan Y, Fibich G and Weinstein M I 2006 *Phys. Rev. Lett.* **97** 193902
[38] Belmonte-Beitia J, Pérez-García V M, Vekslerchik V and Torres P J 2007 *Phys. Rev. Lett.* **98** 064102
[39] Abdullaev F K, Gammal A, Salerno M and Tomio L 2008 *Phys. Rev. A* **77** 023615
[40] Kartashov Y V, Malomed B A, Vysloukh V A and Torner L 2009 *Opt. Lett.* **34** 770
　　Kartashov Y V, Malomed B A, Vysloukh V A and Torner L 2009 *Opt. Lett.* **34** 3625
[41] Zeng J and Malomed B A 2012 *Phys. Rev. A* **85** 023824
[42] Shi J, Zeng J and Malomed B A 2018 *Chaos* **28** 075501
[43] Gao X and Zeng J 2018 *Front. Phys.* **13** 130501
[44] Salasnich L, Cetoli A, Malomed B A, Toigo F and Reatto L 2007 *Phys. Rev. A* **76** 013623
[45] Kartashov Y V, Vysloukh V A and Torner L 2008 *Opt. Lett.* **33** 1747
　　Kartashov Y V, Vysloukh V A and Torner L 2008 *Opt. Lett.* **33** 2173
[46] Sakaguchi H and Malomed B A 2010 *Phys. Rev. A* **81** 013624
[47] Zeng J and Malomed B A 2012 *Phys. Scr.* T **149** 014035
[48] Borovkova O V, Kartashov Y V, Torner L and Malomed B A 2011 *Phys. Rev. E* **84** 035602(R)
[49] Driben R, Kartashov Y V, Malomed B A, Meier T and Torner L 2014 *Phys. Rev. Lett.* **112** 020404
[50] Kartashov Y V, Malomed B A, Shnir Y and Torner L 2014 *Phys. Rev. Lett.* **113** 264101
[51] Zeng J and Malomed B A 2012 *Phys. Rev. E* **86** 036607
[52] Tian Q, Wu L, Zhang Y and Zhang J-F 2012 *Phys. Rev. E* **85** 056603
[53] Cardoso W B, Zeng J, Avelar A T, Bazeia D and Malomed B A 2013 *Phys. Rev. E* **88** 025201
[54] Wu Y, Xie Q, Zhong H, Wen L and Hai W 2013 *Phys. Rev. A* **87** 055801
[55] Driben R, Kartashov Y V, Malomed B A, Meier T and Torner L 2014 *New J. Phys.* **16** 063035
[56] Zeng J and Malomed B A 2017 *Phys. Rev. E* **95** 052214
[57] Kartashov Y V, Vysloukh V A and Torner L 2006 *Phys. Rev. Lett.* **96** 073901
[58] Suntsov S, Makris K G, Christodoulides D N, Stegeman G I, Haché A, Morandotti R, Yang H, Salamo G and Sorel M 2006 *Phys. Rev. Lett.* **96** 063901
[59] Rosberg C R, Neshev D N, Krolikowski W, Mitchell A, Vicencio R A, Molina M I and Kivshar Y S 2006 *Phys. Rev. Lett.* **97** 083901
[60] Szameit A, Kartashov Y V, Dreisow F, Pertsch T, Nolte S, Tünnermann A and Torner L 2007 *Phys. Rev. Lett.* **98** 173903
[61] Dong L and Li H 2010 *J. Opt. Soc. Am. B* **27** 1179
[62] Ye F, Kartashov Y V and Torner L 2006 *Phys. Rev. A* **74** 063616
[63] Aceves A B, Moloney J V and Newell A C 1989 *Phys. Rev. A* **39** 1809
[64] Aceves A B, Moloney J V and Newell A C 1989 *Phys. Rev. A* **39** 1828
[65] Louis P J Y, Ostrovskaya E A, Savage C M and Kivshar Y S 2003 *Phys. Rev. A* **67** 013602
[66] Efremidis N K and Christodoulides D N 2003 *Phys. Rev. A* **67** 063608